# Strengthening the magnetic interactions in pseudobinary first-row transition metal thiocyanates, M(NCS)$_2$

Euan N. Bassey,[1] Joseph A. M. Paddison,[2,3,4] Evan N. Keyzer,[1] Jeongjae Lee,[1,5] Pascal Manuel,[6] Ivan da Silva,[6] Siân E. Dutton,[3] Clare P. Grey[1,*] and Matthew J. Cliffe[1,7,*]

[1]Department of Chemistry, Lensfield Road, University of Cambridge, CB2 1EW, United Kingdom

[2]Churchill College, University of Cambridge, Storey's Way, Cambridge, CB3 0DS, United Kingdom

[3]Cavendish Laboratory, Department of Physics, University of Cambridge, JJ Thompson Avenue, Cambridge, CB3 0HE, United Kingdom

[4]Materials Science & Technology Division, Oak Ridge National Laboratory, Oak Ridge, TN 37831, United States of America

[5]School of Earth and Environmental Sciences, Seoul National University, Seoul 08826, Korea

[6]ISIS Facility, STFC Rutherford Appleton Laboratory, Harwell Oxford, Didcot, OX11 0QX, United Kingdom

[7]School of Chemistry, University Park, Nottingham, NG7 2RD, United Kingdom

**ABSTRACT:** Understanding the effect of chemical composition on the strength of magnetic interactions is key to the design of magnets with high operating temperatures. The magnetic divalent first-row transition metal (TM) thiocyanates are a class of chemically simple layered molecular frameworks. Here, we report two new members of the family, manganese (II) thiocyanate, Mn(NCS)$_2$, and iron (II) thiocyanate, Fe(NCS)$_2$. Using magnetic susceptibility measurements on these materials and on cobalt (II) thiocyanate and nickel (II) thiocyanate, Co(NCS)$_2$ and Ni(NCS)$_2$, respectively, we identify significantly stronger net antiferromagnetic interactions between the earlier TM ions—a decrease in the Weiss constant, $\theta$, from 29 K for Ni(NCS)$_2$ to −115 K for Mn(NCS)$_2$—a consequence of more diffuse 3$d$ orbitals, increased orbital overlap and increasing numbers of unpaired $t_{2g}$ electrons. We elucidate the magnetic structures of these materials: Mn(NCS)$_2$, Fe(NCS)$_2$ and Co(NCS)$_2$ order into the same antiferromagnetic commensurate ground state, whilst Ni(NCS)$_2$ adopts a ground state structure consisting of ferromagnetically ordered layers stacked antiferromagnetically. We show that significantly stronger exchange interactions can be realised in these thiocyanate frameworks by using earlier TMs.

## 1. INTRODUCTION

The rational design and synthesis of new magnetic materials tailored to particular functions requires an understanding of the fundamental interactions taking place between magnetic centres. Magnetic molecular framework materials—that is, systems in which magnetic centres are connected *via* molecular ligands into lattices—present an excellent opportunity to study these interactions and their chemical origins.

Using molecular bridging ligands to connect paramagnetic metal centres can produce open and flexible structures, permitting potential applications such as multiferroics,[1–3] magnetostrictive materials[4] and magnetic sensors.[5–7] The properties of molecular frameworks are not determined by long-range electrostatic forces—as in oxide frameworks—but are instead dominated by short-range coordination bonds between the metal centre, *M* and its ligands, *L*.[8] The nature and delocalisation of the *M*–*L* bond controls the exchange interactions and therefore the magnetic properties of the material. By careful choice of *L*, very anisotropic structures can be created, which may have low-dimensional magnetic properties.[9–15]

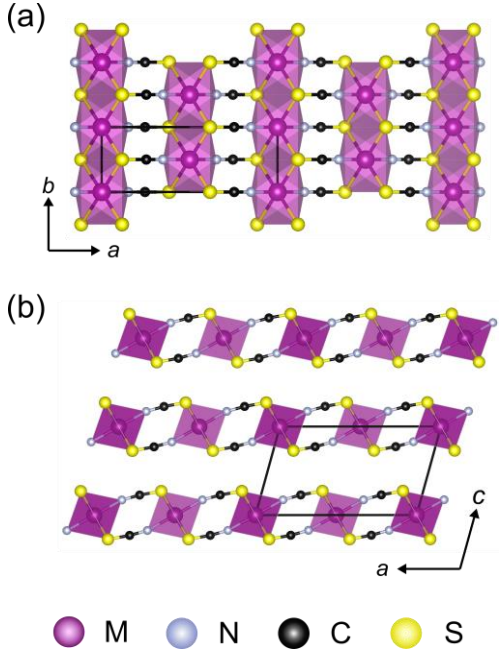

**Figure 1. (a)** Projection of a (2×2×1) supercell of $M(NCS)_2$ ($M$ = Mn, Fe, Co, Ni) down the $c$ axis, highlighting the layers of edge-sharing [$M^{2+}N_2S_4$] octahedra. **(b)** Projection of a (2×1×2) supercell of $M(NCS)_2$ down the $b$ axis, showing the stacking of layers.

Low-dimensional magnetic materials are an example of many-body quantum systems amenable to exact mathematical treatment.[16] In particular, the study of the behaviour of low-dimensional magnets is of critical importance to spin-liquids[17,18] and high-temperature superconductivity.[19]

Thiocyanate, (NCS)⁻, is a promising ligand in magnetic molecular frameworks, as it is capable of promoting strong superexchange interactions between paramagnetic metal centres—for example, $J$ = +230 K in $Cu_2(NCS)_4(bpm)$ (bpm = 2,2-bipyrimidine).[9,20–25] However, the majority of the frameworks studied thus far incorporate ancillary ligands, which can decrease metal-metal connectivity in the framework and thereby the net strength of interactions (as measured by the Weiss constant, $\theta$).[20,21]

Despite the range of complex magnetic thiocyanate compounds reported,[26] the parent pseudobinary system, $M(NCS)_2$, is relatively unexplored, with only three known magnetic examples: $Co(NCS)_2$,[20] $Ni(NCS)_2$[27,28] and $Cu(NCS)_2$.[9] $Cu(NCS)_2$ is a quasi-one-dimensional magnet, due to the Jahn-Teller distortion of the $Cu^{2+}$ ions, with strong superexchange interactions along the $Cu(NCS)_2$ chain ($J$ = 133 K), but a significantly lower Néel temperature, $T_N$ = 12 K.[9]

Previous work has established that both $Co(NCS)_2$[20] and $Ni(NCS)_2$[28] order antiferromagnetically, with $T_N$ = 22 K and 52 K, respectively. However, the value of $\theta$ for $Ni(NCS)_2$ suggested net ferromagnetic interactions ($\theta$ = 39.8 K);[28] in contrast, $Co(NCS)_2$ has net antiferromagnetic interactions ($\theta$ = −40 K).[20] At present, the magnetic structures of $Co(NCS)_2$ and $Ni(NCS)_2$ are unknown, which makes rationalising the differences in the values of $\theta$ challenging.

The structures adopted by $M(NCS)_2$ are directly analogous to the corresponding transition metal (TM) halides, $MX_2$ [Figure 1]. $MX_2$ consist of layers of edge-sharing $M^{2+}$ octahedra with weak van der Waals interactions between the layers [Figure 1].[20,27] The triangular metal sublattice lends itself to geometric frustration,[29–31] and hence leads to unusual magnetic properties, such as multiferroic behaviour and helimagnetism.[32–36] Furthermore, the strength of magnetic interactions in $MX_2$ is comparable to those in $M(NCS)_2$, for example $Ni(NCS)_2$ ($T_N$ = 52 K)[28] and $NiBr_2$ ($T_N$ = 44 K).[32,37]

In $M(NCS)_2$ ($M$ = Co, Ni), the metal sites form layers of an anisotropic triangular lattice, and so $M(NCS)_2$ have the potential to show similar unusual magnetic behaviour. The two nearest-neighbour interactions within the layers are $J_1$, along the $M$–S–$M$ chains (along the [010] direction), and $J_2$, through $M$–NCS–$M$ linkages [Figure 2(b) and (c)]; the Heisenberg Hamiltonian for this system may be written as:

$$\hat{H} = \sum_{i,j} \mathbf{S}_i J_{ij} \mathbf{S}_j, \quad (1)$$

where the summation is taken over all nearest-neighbour pairs of spins $i$ and $j$ with a Heisenberg exchange constant $J_{ij}$ between them. Here, each pair is counted once and positive values of $J_{ij}$ correspond to antiferromagnetic interactions. In $Cu(NCS)_2$, just as in $CuCl_2$ and $CuBr_2$, ordering of the Jahn-Teller distortion lowers the structural symmetry and disrupts the superexchange pathways, producing quasi one-dimensional magnetism.[9,38,39]

The spatially anisotropic triangular lattice may be characterised using a single parameter, $\phi = \tan^{-1}\left(\frac{J_1}{J_2}\right)$ [Figure 2(a)].[40,41] By varying $J_1$ and

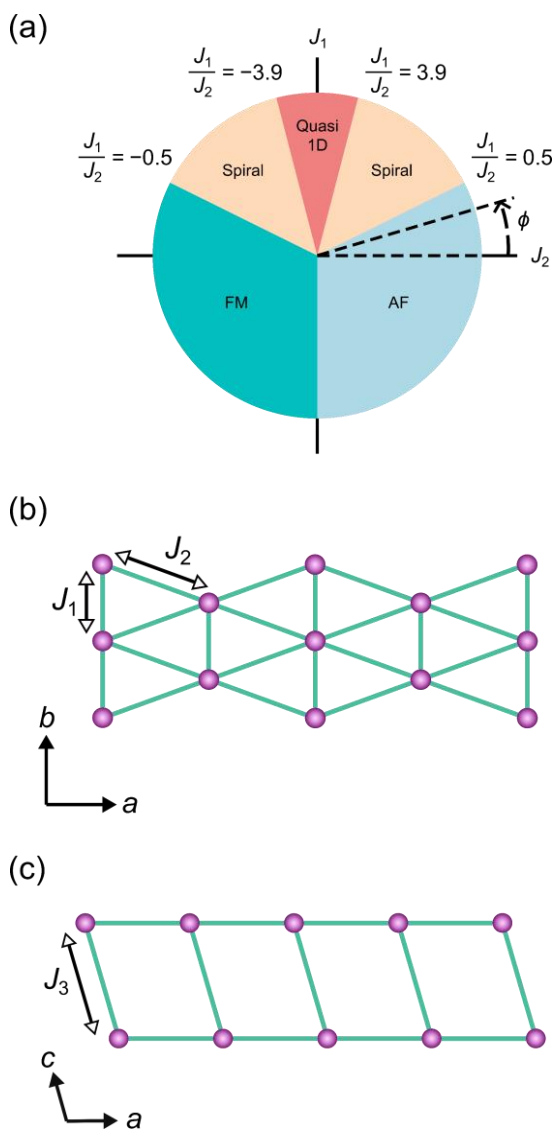

**Figure 2. (a)** Classical Heisenberg magnetic phase diagram for an anisotropic two-dimensional triangular lattice, adapted from Ref. 41. The nearest-neighbour exchange interactions $J_1$ and $J_2$, are illustrated in **(b)**, whilst the interlayer coupling constant, $J_3$, is illustrated in **(c)**.

$J_2$, the isotropic triangular lattice ($J_1 = J_2$, $\phi = \frac{\pi}{4}$) can be transformed into a square lattice ($J_1 = 0$, $\phi = 0, \pi$), or a quasi-one-dimensional chain ($J_2 = 0$, $\phi = \frac{\pi}{2}$). Although we anticipate the presence of a small finite interlayer coupling, perhaps mediated via S⋯S interlayer contacts[42] ($J_3$, where we expect $J_3 \ll J_1, J_2$), this two-dimensional phase diagram will still provide a qualitative reference for the observed phases, as has been found for other van der Waals magnets with comparable interlayer distances.[32–35,41,43–45]

In this work, we synthesised and characterised $M(NCS)_2$ ($M$ = Mn, Fe, Co and Ni)—$M$ = Mn and Fe for the first time. The materials were characterised using powder X-ray diffraction, thermogravimetric analysis and diffuse reflectance spectroscopy. In addition, we measured their magnetic properties using magnetic susceptibility measurements and powder neutron diffraction (PND) and determined constraints on the relative sizes of their magnetic exchange interactions.

We show that the net strength of superexchange interactions, as measured by the Weiss constant, $\theta$, increases and becomes increasingly antiferromagnetic as we move to earlier first-row TM cations.

From PND measurements, $Mn(NCS)_2$, $Fe(NCS)_2$ and $Co(NCS)_2$ are observed to adopt the same commensurate stripe-ordered magnetic ground state with ordering vector $\boldsymbol{k} = [100]^*$. In contrast, $Ni(NCS)_2$ adopts a ground state magnetic structure with ordering vector $\boldsymbol{k} = \left[00\tfrac{1}{2}\right]^*$, consistent with its very different (and positive) Weiss constant.

## 2. MATERIALS AND METHODS

### 2.1. Synthesis

The synthetic procedures for each member of the $M(NCS)_2$ ($M$ = Mn, Fe, Co and Ni) used were broadly similar. We therefore provide a general synthetic route for $M(NCS)_2$ here; complete synthetic routes for each compound are given in the supporting information.

For $M$ = Mn, Co and Ni, TM sulphate salts were dissolved in the minimum volume of deionised $H_2O$ and added to a saturated solution of $Ba(SCN)_2 \cdot 3H_2O$. For $M$ = Fe, a solution of KSCN in dry acetonitrile was added to $Fe(BF_4)_2 \cdot 6H_2O$. In all cases, a white precipitate (coloured by the strongly coloured solution) formed immediately and the reaction mixture was stirred in air ($M$ = Mn, Co and Ni) or under a nitrogen atmosphere ($M$ = Fe). The solvent was then removed in vacuo to generate a microcrystalline powder.

**2.2. Powder X-Ray Diffraction (PXRD)** Phase purity was assessed via powder diffraction measurements on a PANalytical Empyrean Diffractometer using Cu-K$\alpha$ radiation ($\lambda$ = 1.541 Å) in Bragg-Brentano geometry. Diffraction patterns were recorded over the range $2\theta$ = 5–80° using a step size of 0.02° and a scan speed of 0.01° s$^{-1}$. Due to their sensitivity to moisture and air, the diffraction patterns of Mn(NCS)$_2$ and Fe(NCS)$_2$ were measured by encasing the samples between polyimide (Kapton) films. All diffraction patterns were analysed via Pawley[46] and Rietveld[47,48] refinements using TOPAS Academic 6 structure refinement software.[49,50]

### 2.3. Diffuse Reflectance Spectroscopy

Diffuse reflectance spectra were recorded on an Agilent Technologies UV-VIS spectrometer, connected *via* optical fibre to a Cary 50 Diffuse Reflectance Accessory, using a wavelength range $\lambda$ = 200–1000 nm, with step size 1.00 nm and scan rate of 10 nm s$^{-1}$. In all cases, samples were diluted with BaSO$_4$ powder—either in a 1:10 mass ratio (Fe(NCS)$_2$, Co(NCS)$_2$ and Ni(NCS)$_2$) or a 1:1 mass ratio (Mn(NCS)$_2$)—and the mixture ground to produce a homogeneous powder, which was then loaded between two quartz discs. For Mn(NCS)$_2$ and Fe(NCS)$_2$, the homogeneous powders were prepared inside an Ar-filled glovebox and the quartz discs sealed with Parafilm; the spectra for these compounds were acquired within half an hour of removing the samples and discs from the glovebox.

For all materials, the spectra were averaged over multiple measurements; spikes in the average due to erroneous spikes in the raw data—i.e. spikes in one spectrum which do not repeat in the other spectra, likely due to specular reflection from the powder—were removed from the average and the 'spiked' data point replaced with the average intensity either side of the spike.

### 2.4. Thermogravimetric Analysis (TGA)

Thermogravimetric data for each compound were recorded with a Mettler-Toledo Thermogravimetric Analysis/Simultaneous Differential Thermal Analysis (TGA/SDTA) 851 Thermobalance. Each powder sample (20–50 mg) was loaded into an alumina crucible and heated from 50°C to 600°C at a heating rate of 10°C min$^{-1}$ under a nitrogen atmosphere. The data collected was measured relative to a background blank TGA curve, recorded using the same alumina crucible, temperature range and heating rate, under a nitrogen atmosphere.

### 2.5. Magnetic Susceptibility Measurements

The magnetic susceptibility measurements were carried out on powder samples (10–20 mg) using a Quantum Design Magnetic Property Measurement System 3 (MPMS) superconducting quantum interference device (SQUID) magnetometer. The zero-field cooled (ZFC) and field-cooled (FC) susceptibilities were measured in a field of 0.01 T over a temperature range 2–300 K. As $M(H)$ is linear in this field range, the small-field approximation to the susceptibility, $\chi \simeq \frac{M}{H}$, was assumed to be valid. The data for each compound were corrected for diamagnetism of the sample using Pascal's constants.[51]

### 2.6. Powder Neutron Diffraction (PND)

Powder neutron diffraction measurements were carried out at the ISIS Pulsed Neutron and Muon Source using the WISH (Mn(NCS)$_2$, Co(NCS)$_2$, Ni(NCS)$_2$) and GEM (Fe(NCS)$_2$) instruments.[52] Samples of Mn(NCS)$_2$ (4.76 g), Fe(NCS)$_2$ (2.26 g), Co(NCS)$_2$ (2.44 g) and Ni(NCS)$_2$ (4.76 g) were loaded into thin-walled vanadium canisters. The canister diameters were 11 mm for Mn(NCS)$_2$ and Ni(NCS)$_2$; 6 mm for Co(NCS)$_2$ and 6 mm with an indium seal for Fe(NCS)$_2$. Each sample was loaded to a height of at least 40 mm, to ensure the full beam illuminates the sample.

Each sample was first cooled to the base temperature (1.5 K for Mn(NCS)$_2$, Co(NCS)$_2$ and Ni(NCS)$_2$ and 10 K for Fe(NCS)$_2$) and diffraction patterns then collected at a series of temperatures through $T_N$. The complete list of temperature steps and data collections may be found in the supporting information. The data were corrected for absorption effects using the Mantid software package.[53]

For each compound, the nuclear structure was determined by Rietveld refinement against powder neutron diffraction data collected above $T_N$, using a model derived from the previously reported single crystal structure of Ni(NCS)$_2$.[27] All refinements were carried out using TOPAS Academic 6.0.[50]

Rietveld refinements using the candidate magnetic irreducible representations (irreps) were carried out for each compound separately, which showed that in each case only one of the two single irrep structures was consistent with the experimental data. Including the second irrep did not significantly improve the fit to the data. On this basis, we refined the magnetic structures using only the $mY_2^+$ irrep for Mn(NCS)$_2$, Co(NCS)$_2$ and Fe(NCS)$_2$, and the $mA_1^+$ irrep for Ni(NCS)$_2$. All refinements were carried out by simultaneously refining against data collected on multiple banks of detectors: on WISH, for Mn(NCS)$_2$, banks 2–5; for Co(NCS)$_2$, banks 1–5 and for Ni(NCS)$_2$, banks 2–5; on GEM, for Fe(NCS)$_2$, banks 2–5.

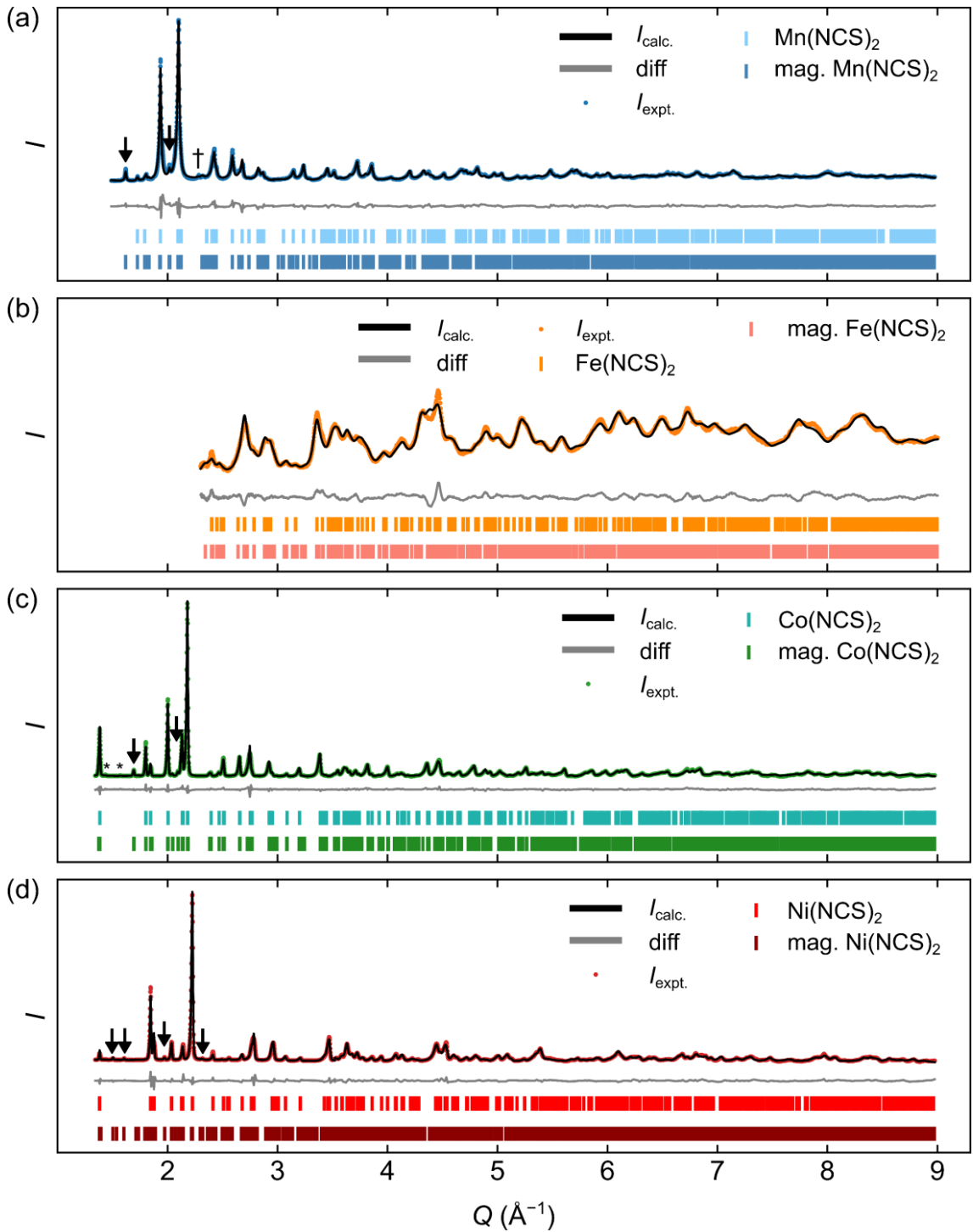

**Figure 3.** Rietveld refinements of the neutron powder diffraction patterns collected at 1.5 K for Mn(NCS)$_2$, **(a)**; at 1.5 K for Co(NCS)$_2$, **(c)** and 1.5 K for Ni(NCS)$_2$ **(d)**, all on Bank 5 of WISH. **(b)** shows the pattern collected at 10 K for Fe(NCS)$_2$ on Bank 5 of GEM. Arrows lie directly above magnetic peaks, asterisks are directly above magnetic reflections from impurity phases, and daggers are directly above non-magnetic reflections from impurity phases. $R_{wp}$ = 7.54% for Mn(NCS)$_2$, 2.99% for Fe(NCS)$_2$, 4.06% for Co(NCS)$_2$ and 6.25% for Ni(NCS)$_2$. The low-$Q$ regions of the diffraction patterns are shown in more detail in Figure 8, to emphasize the magnetic Bragg peaks.

In all cases, the background was fit with a 12-term Chebyshev polynomial. For the final refinements of the data collected for Mn(NCS)$_2$ using the WISH diffractometer and for the final refinement of the data collected for Fe(NCS)$_2$ on the GEM diffractometer, it proved necessary to refine the Voigt peak-shape parameters separately for high $Q$ and low $Q$ data, due to their unusual $Q$-dependence.

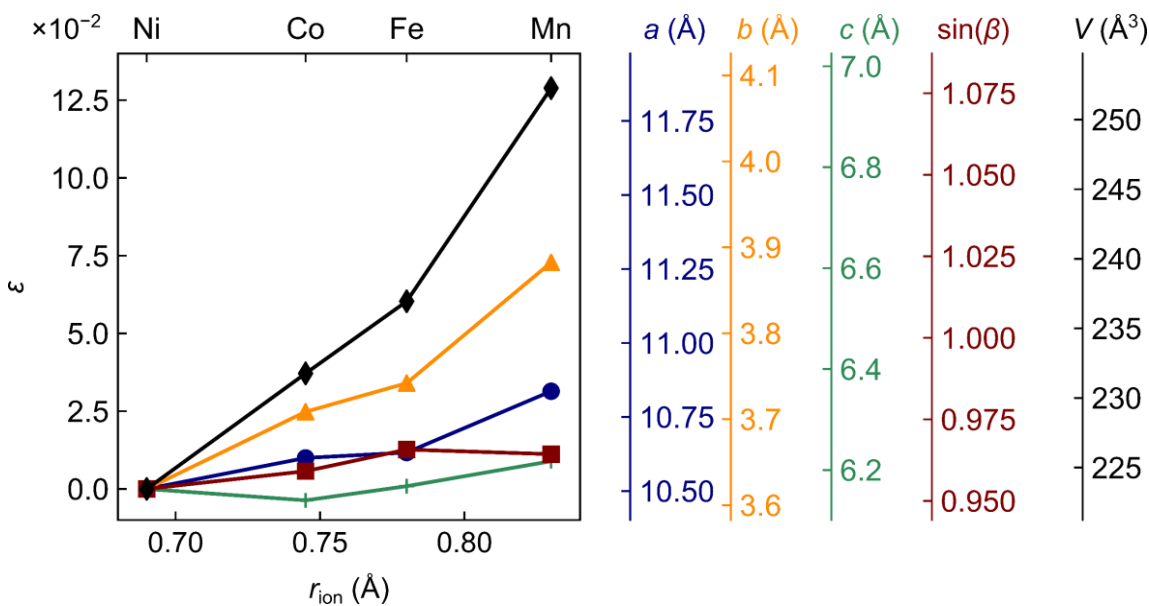

**Figure 4.** Variation in the strain of Rietveld-derived PND lattice parameters and nuclear unit cell volume (at the base temperature of PND results) with the ionic radius of $M^{2+}$, $r_{ion}$. The strain in each parameter, $x$ ($x = a, b, c,$ $\sin(\beta), V$), is calculated via $\varepsilon = \frac{x - x_{Ni}}{x_{Ni}}$ – i.e. the strain is calculated relative to the parameters obtained for Ni(NCS)$_2$. The corresponding values of the lattice parameters are shown on the right-hand axes.

For all refinements, the lattice parameters, atomic positions and the magnitudes and directions of the magnetic moments were allowed to refine freely, aside from restraints on the C–N (*ca.* 1.15 Å) and C–S (*ca.* 1.65 Å) bond lengths. For Mn(NCS)$_2$, Co(NCS)$_2$ and Ni(NCS)$_2$, the same set of freely refining anisotropic atomic displacement parameters was used for each atom, whilst the same isotropic atomic displacement parameter was refined for each atom in Fe(NCS)$_2$ [Table S6]. The bond lengths and angles were consistent with those expected from previous studies [Table S1].[20,27]

## 3. RESULTS

**3.1. Bulk characterisation** The $M$(NCS)$_2$ family members ($M$ = Mn$^{2+}$, Fe$^{2+}$, Co$^{2+}$, Ni$^{2+}$) were synthesised via salt metathesis reactions, driven by precipitation of an insoluble side-product (BaSO$_4$ for $M$ = Mn, Co, Ni; KBF$_4$ for $M$ = Fe). Apart from Ni(NCS)$_2$, all compounds crystallised as solvates; the co-crystallised solvent was removed by heating either *in vacuo* ($M$ = Fe, Mn) or in air ($M$ = Co). Whilst Co(NCS)$_2$ and Ni(NCS)$_2$ were stable in air, Mn(NCS)$_2$ and Fe(NCS)$_2$ were moisture- and air-sensitive, respectively. The phase purity of all materials was checked initially using PXRD [Figure S1], revealing the presence of trace quantities (< 1 wt. %) of impurities such as unreacted starting materials or hydrates. To confirm whether any solvent remained trapped in the synthesised frameworks, we carried out thermogravimetric analysis (TGA) [Figure S2], which revealed minimal quantities of water lost from each material.

Quantitative Rietveld refinements of the high-$Q$ powder neutron diffraction (PND) data collected confirmed that all four compounds were isostructural [Figure 3], crystallising in the space group $C2/m$, as anticipated for the similar chemistries and ionic radii of the divalent first row TM cations.[54]

The lattice parameters and atomic coordinates derived from Rietveld refinements are shown in Table 1. The lattice parameters and unit cell volume vary approximately linearly with cationic radius, $r_{ion}$ [Figure 4]. The $b$ lattice parameter depends only on the $M$–S bond lengths and $M$–S–$M$ bond angle and therefore changes proportionately the most as $M$ varies. The observed changes in $b$ are also consistent with the expected differences in $M$–S bond lengths across the TM series.[21] The $a$ lattice parameter

**Table 1.** Rietveld-derived lattice parameters and atomic coordinates, based on the powder neutron diffraction data collected at 1.5 K for Mn(NCS)$_2$, Co(NCS)$_2$ and Ni(NCS)$_2$ and at 10 K for Fe(NCS)$_2$. The space group for all compounds is $C2/m$.

**Mn(NCS)$_2$**

| | $a$ (Å) | 10.8370(17) | $\alpha$ (°) | 90 |
|---|---|---|---|---|
| | $b$ (Å) | 3.8824(6) | $\beta$ (°) | 105.348(2) |
| | $c$ (Å) | 6.2175(9) | $\gamma$ (°) | 90 |

| | Site | $x$ | $y$ | $z$ |
|---|---|---|---|---|
| Mn | 2a | 0 | 0 | 0 |
| N | 4i | −0.1445(2) | 0 | 0.1723(4) |
| C | 4i | −0.2465(4) | 0 | 0.2278(6) |
| S | 4i | −0.3743(7) | 0 | 0.2721(11) |

**Fe(NCS)$_2$**

| | $a$ (Å) | 10.630(16) | $\alpha$ (°) | 90 |
|---|---|---|---|---|
| | $b$ (Å) | 3.742(6) | $\beta$ (°) | 105.038(7) |
| | $c$ (Å) | 6.168(10) | $\gamma$ (°) | 90 |

| | Site | $x$ | $y$ | $z$ |
|---|---|---|---|---|
| Fe | 2a | 0 | 0 | 0 |
| N | 4i | −0.14037(13) | 0 | 0.17658(18) |
| C | 4i | −0.24139(14) | 0 | 0.2220(2) |
| S | 4i | −0.3897(3) | 0 | 0.2668(4) |

**Co(NCS)$_2$**

| | $a$ (Å) | 10.6118(4) | $\alpha$ (°) | 90 |
|---|---|---|---|---|
| | $b$ (Å) | 3.70869(11) | $\beta$ (°) | 106.4401(8) |
| | $c$ (Å) | 6.13996(19) | $\gamma$ (°) | 90 |

| | Site | $x$ | $y$ | $z$ |
|---|---|---|---|---|
| Co | 2a | 0 | 0 | 0 |
| N | 4i | −0.13799(7) | 0 | 0.17056(13) |
| C | 4i | −0.23820(11) | 0 | 0.2171(2) |
| S | 4i | −0.3782(2) | 0 | 0.2610(4) |

**Ni(NCS)$_2$**

| | $a$ (Å) | 10.5070(5) | $\alpha$ (°) | 90 |
|---|---|---|---|---|
| | $b$ (Å) | 3.61889(5) | $\beta$ (°) | 107.509(3) |
| | $c$ (Å) | 6.16252(16) | $\gamma$ (°) | 90 |

| | Site | $x$ | $y$ | $z$ |
|---|---|---|---|---|
| Ni | 2a | 0 | 0 | 0 |
| N | 4i | −0.1325(2) | 0 | 0.1699(4) |
| C | 4i | −0.2367(3) | 0 | 0.2086(7) |
| S | 4i | −0.3778(6) | 0 | 0.2543(9) |

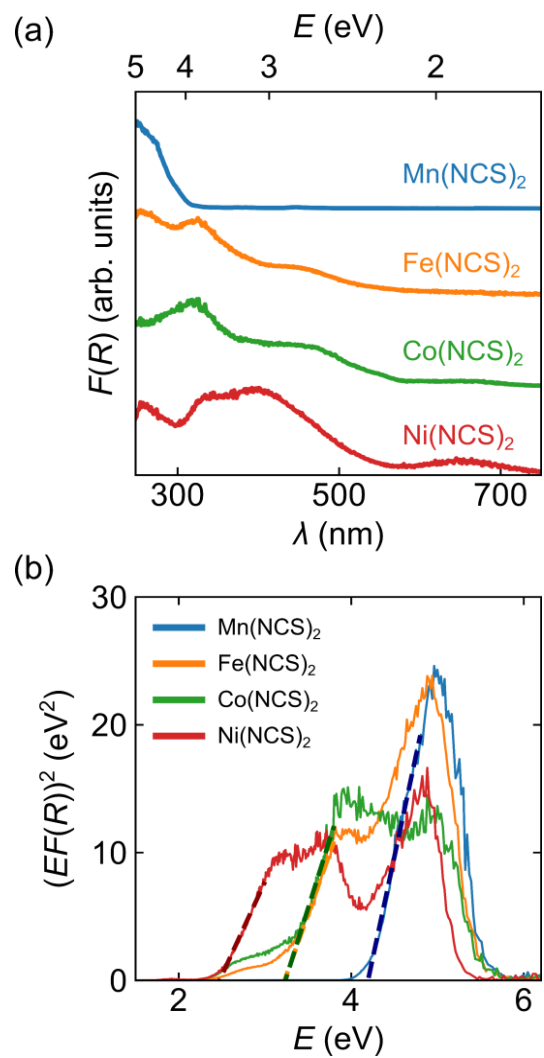

**Figure 5. (a)** Diffuse reflectance spectra for $M$(NCS)$_2$ ($M$ = Mn, Fe, Co, Ni). **(b)** Tauc plots of the diffuse reflectance data: extrapolation of the data to $(F(R)h\nu)^2 = 0$ (indicated by the dashed line) yields the band gap for each material.

is dominated by the length of (NCS)$^-$ anion (which lies along the $a$ axis), rather than the $M$–N and $M$–S bonds (which are oriented at an angle to the $a$ direction) [Figure 1]. Since the length of (NCS)$^-$ remains approximately constant regardless of the identity of $M$, changes in $M$ have a small effect on the size of $a$. The $c$ lattice parameter remains approximately constant across the series and is determined by the interlayer van der Waals interactions, suggesting these interactions have similar strengths across the first-row TM series.

All compounds except Mn(NCS)$_2$ were strongly coloured microcrystalline powders: Fe(NCS)$_2$ was orange-brown, Co(NCS)$_2$ was red-brown[20] and Ni(NCS)$_2$ green-brown; Mn(NCS)$_2$ was pale yellow.[27] To quantitatively assess the variation in observed colours across the $M$(NCS)$_2$ series, we recorded diffuse reflectance UV-Vis spectra [Figure 5(a)].

The observed intense transitions correspond to ligand-to-metal charge transfer (LMCT) transitions: from states dominated by (NCS)$^-$–

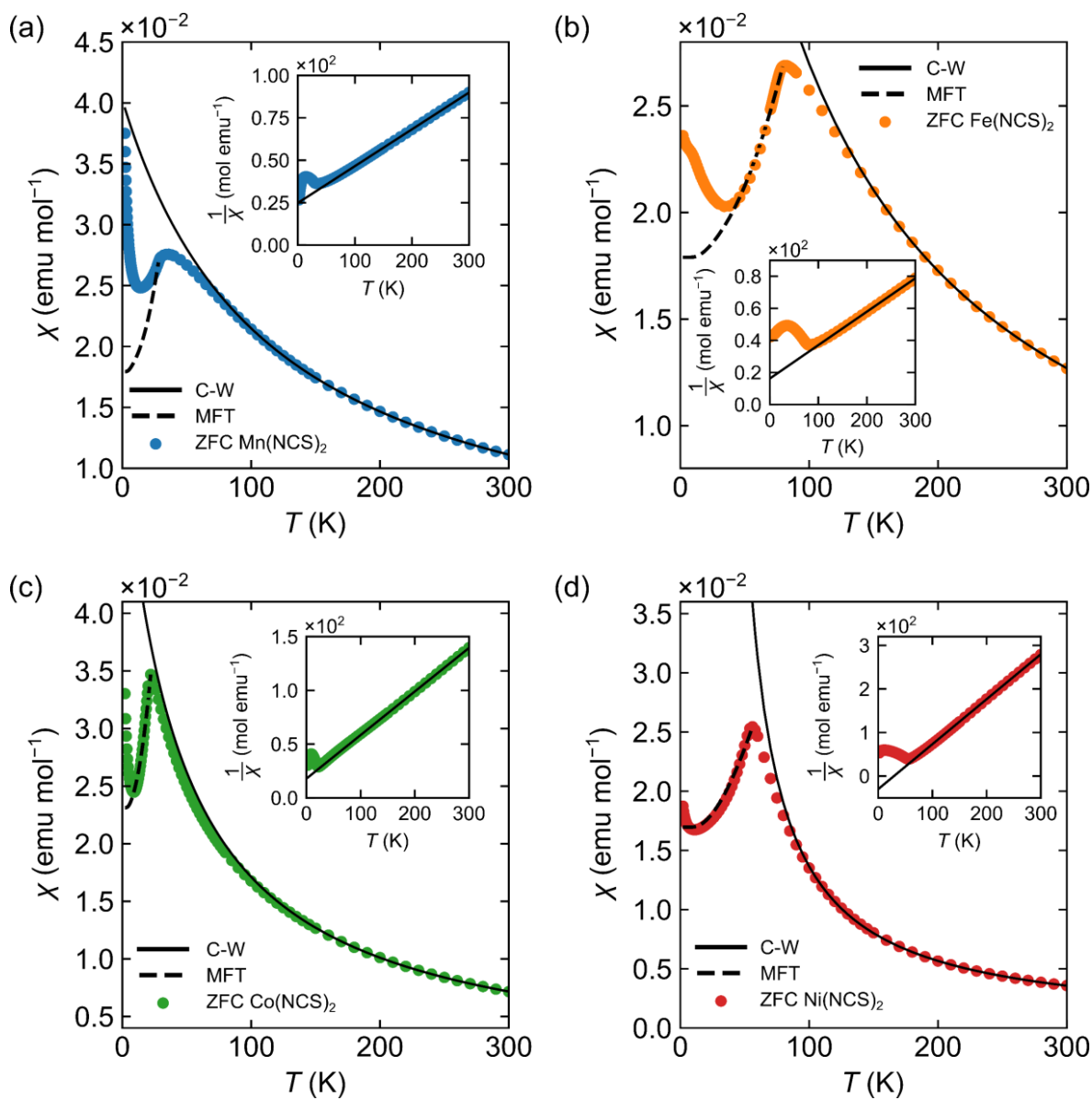

**Figure 6.** Zero-field cooled (ZFC) magnetic susceptibility data collected for $Mn(NCS)_2$ (**a**), $Fe(NCS)_2$ (**b**), $Co(NCS)_2$ (**c**) and $Ni(NCS)_2$ (**d**) in a constant magnetic field strength $H$ = 0.01 T. The Curie-Weiss law was used to model the high-temperature ($T$ > 150 K) data, while a low-temperature mean-field theory (MFT) model was used for the data $T < T_N$. Insets show the inverse of magnetic susceptibility, used to examine the Curie-Weiss law fit.

based orbitals to states dominated by the metal 3$d$ orbitals.[55–57] The additional weak absorption bands observed for $Fe(NCS)_2$, $Co(NCS)_2$ and $Ni(NCS)_2$ likely correspond to $d$-$d$ transitions.[58] The optical (indirect) band gaps were extracted using Tauc fits to the data [Figure 5(b)],[59] giving the following values: 4.2(1) eV for $Mn(NCS)_2$, 3.2(1) eV for $Fe(NCS)_2$, 3.2(1) eV for $Co(NCS)_2$ and 2.5(1) eV for $Ni(NCS)_2$. On moving to later TM ions, the $d$ orbitals decrease in energy and the crystal field splitting increases, resulting in a lower energy (longer wavelength) transition and smaller band gap. This trend is consistent with the small observed band gap for $Cu(NCS)_2$ (1.3 eV), which is lowered further due to Jahn-Teller distortions of the $Cu^{2+}$ ions.[9]

**3.2. Bulk magnetic measurements** To assess the change in the magnetic properties as the identity of $M$ in $M(NCS)_2$ varies, we next went on to measure the bulk magnetic susceptibilities of these compounds at $H$ = 0.01 T [Figure 6]. In addition, isothermal magnetisation measurements ($M(H)$ curves) were carried out [Figures S3–S6]; these showed that, in the region −7 to +7 T, saturation is not achieved.

**Table 2.** Bulk magnetic susceptibility parameters, extracted from the raw magnetic susceptibility data ($T_N$), high-temperature Curie-Weiss law fits ($\theta$, $\mu_{eff}$ and $g$) and Rietveld refinements of the low-temperature PND data (staggered moments, $m_{sta.}$). Experimental standard errors are given in parentheses.

|  | Mn(NCS)$_2$ | Fe(NCS)$_2$ | Co(NCS)$_2$ | Ni(NCS)$_2$ |
|---|---|---|---|---|
| $\theta$ (K) | −115(3) | −78(3) | −44(1) | +29(1) |
| $T_N$ (K) | 28.0(3) | 78.4(3) | 20.0(5) | 54(2) |
| $\mu_{eff}$ ($\mu_B$) | 6.4(5) | 6.2(5) | 4.4(5) | 2.8(5) |
| $g$ | 2.2(2) | 2.5(2) | 2.3(2) | 2.0(2) |
| $m_{sta.}$ ($\mu_B$) | 4.02(4) | 4.95(6) | 3.02(2) | 1.75(5) |

All four compounds showed evidence of three-dimensional antiferromagnetic ordering [Figure 6]: a sharp change in $\frac{d\chi}{dT}$ at $T_N$ and a rapid decrease in $\chi$ at low temperatures; in all cases, the ZFC and FC susceptibilities did not diverge [Figure S7]. Each compound also shows a rise in the susceptibility at low temperatures ($T$ < 15 K for $M$ = Fe, Co, Ni and $T$ < 35 K for $M$ = Mn), due to small amounts (< 1 wt.%) of paramagnetic impurities and defects, as observed in several other magnetic molecular frameworks.[9,21,60]

For each compound, the high-temperature magnetic susceptibility ($T$ > 150 K) data were fitted to the Curie-Weiss law, yielding values of the Weiss constant, $\theta$, and Curie constant, $C$ [Table 2]. The data collected for Ni(NCS)$_2$ and Co(NCS)$_2$ are broadly consistent with previously reported measurements.[20,28] The presence of a significant residual field (< 20 Oe) was identified from the isothermal magnetisation data [Figures S4–S7]. We have corrected for this field, but it did introduce a significant additional uncertainty. The presence of small quantities of paramagnetic impurities, single-ion anisotropy and the fact that by necessity we are carrying out a Curie-Weiss fit at $T$ < 3|$\theta$| (for Fe(NCS)$_2$ and Mn(NCS)$_2$) will also introduce additional small, systematic errors in the fitted parameters.

The large deviations in $\mu_{eff.}$ from the spin-only values for Fe(NCS)$_2$ and Co(NCS)$_2$ likely arise from spin-orbit coupling, due to the residual orbital angular momentum in the $^5T_{2g}$ and $^4T_{1g}$ terms for high-spin octahedral Fe$^{2+}$ and Co$^{2+}$, respectively.[61–64] The observed deviations for Mn(NCS)$_2$ and Ni(NCS)$_2$ are much smaller, as no first-order orbital contribution is expected.[65,66]

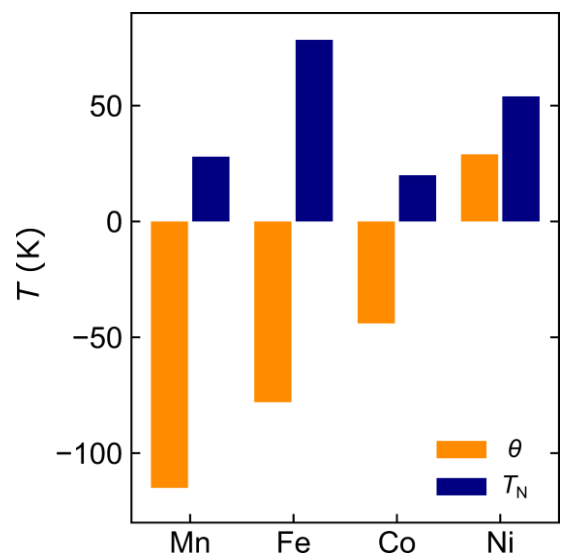

**Figure 7.** Variation of the Weiss temperature, $\theta$, and Néel temperature, $T_N$, across the first-row transition metals, $M$, in the $M$(NCS)$_2$ family.

We note that net magnetic interactions, as measured by $\theta$, become increasingly antiferromagnetic on moving to earlier TM ions [Figure 7], likely due to the more diffuse nature of $M^{2+}$ 3$d$ orbitals earlier in the series, enabling better spatial overlap with the (NCS)$^-$ $\sigma$ and $\pi$ frontier orbitals. The trend in Weiss constants also highlights the considerable increase in the net magnetic interaction strength between spins on moving to earlier TM ions, particularly for Mn(NCS)$_2$ and Fe(NCS)$_2$.

As $M$(NCS)$_2$ are layered materials with weak van der Waals interactions between the layers, we anticipated these materials would display low-dimensional magnetic behaviour. The ratio $f = |\theta|/T_N$ was computed for each compound, as this parameter may be used to assess the extent to which long-range order is suppressed by low-dimensionality or spin frustration.[16] For Mn(NCS)$_2$, $f$ = 4.1(1); for Fe(NCS)$_2$, $f$ = 0.99(4); for Co(NCS)$_2$, $f$ = 2.2(1) and for Ni(NCS)$_2$, $f$ = 0.54(3), which do not suggest strong suppression of long-range order.[16]

The observed variation in susceptibility below $T_N$ was modelled using a powder-average mean-field theory (MFT) model with Heisenberg exchange [Figure 6].[67,68] This model qualitatively accounted for the observed magnetic susceptibilities, but could not be used to extract the three nearest neighbour exchange interaction strengths, $J_1$, $J_2$ and $J_3$ [Figure 2(b)], as these parameters are correlated. We then attempted to determine the values of $J_1$, $J_2$ and $J_3$ using a reaction-field model.[69] However, this proved unsuccessful, as strong correlations between the parameters again precluded their reliable determination.

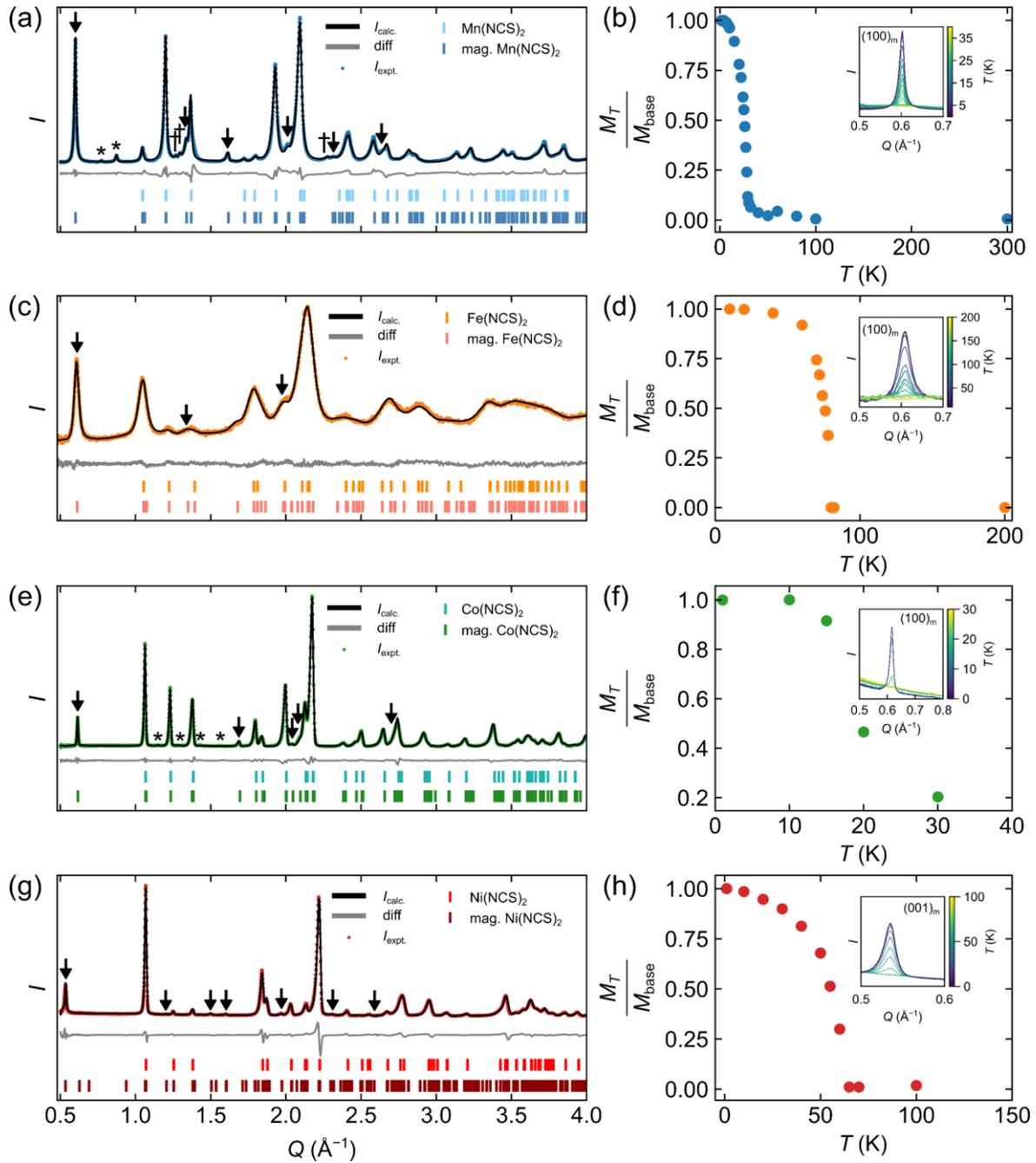

**Figure 8.** Rietveld fits to the PND patterns between $Q = 0.48$ Å$^{-1}$ and $Q = 4.00$ Å$^{-1}$ collected at 1.5 K for Mn(NCS)$_2$ (**a**); at 1.5 K for Co(NCS)$_2$ (**e**); and at 1.5 K for Ni(NCS)$_2$ (**g**), all on Bank 2 of WISH. (**c**) shows the PND pattern for Fe(NCS)$_2$ collected at 10 K on Bank 2 of GEM. Arrows lie directly above magnetic reflections from magnetically ordered $M$(NCS)$_2$; asterisks are directly above reflections from magnetic impurities, whilst daggers are directly above reflections from non-magnetic impurities. $R_{wp}$ = 7.54% for Mn(NCS)$_2$, 2.99% for Fe(NCS)$_2$, 4.06% for Co(NCS)$_2$ and 6.25% for Ni(NCS)$_2$. (**b**), (**d**), (**f**) and (**h**) show how the reduced staggered moment ($M_T/M_{base}$, where $M_T$ is the moment at temperature $T$ and $M_{base}$ the base temperature moment; all moments were obtained from the Rietveld fits) varies as a function of temperature for Mn(NCS)$_2$, Fe(NCS)$_2$, Co(NCS)$_2$ and Ni(NCS)$_2$, respectively. Insets show how the intensities of the magnetic (100) peaks for Mn(NCS)$_2$, Fe(NCS)$_2$ and Co(NCS)$_2$ and the magnetic (001) peak for Ni(NCS)$_2$ vary as a function of temperature.

### 3.3. Magnetic ground state from neutron diffraction experiments

The dominant magnetic interactions in the $M$(NCS)$_2$ frameworks may be understood by examining their ordered magnetic structures. As such, we carried out low-temperature neutron diffraction experiments to determine the magnetic ground state of these frameworks. All four compounds were observed to adopt ordered commensurate ground states.

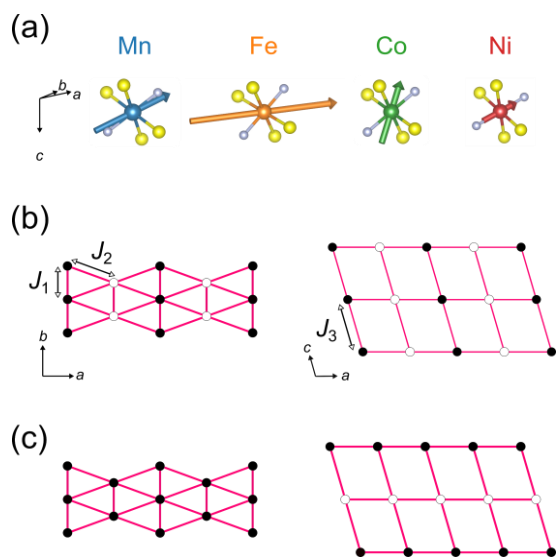

**Figure 9. (a)** Schematic of the ordered magnetic moments of $Mn^{2+}$, $Fe^{2+}$, $Co^{2+}$ and $Ni^{2+}$ in $M(NCS)_2$ (N atoms in grey-blue, S in yellow). The magnetic lattice adopted by $Mn(NCS)_2$, $Fe(NCS)_2$ and $Co(NCS)_2$ is shown in **(b)**, whilst **(c)** shows the magnetic lattice of $Ni(NCS)_2$. Filled and empty circles denote collinear moments with opposite spatial orientations.

On cooling $Mn(NCS)_2$ below $T_N$ (= 28 K), superlattice reflections were observed, corresponding to a propagation vector $\boldsymbol{k} = [100]^*$. Similarly, for $Fe(NCS)_2$ and $Co(NCS)_2$, superlattice reflections corresponding to the $\boldsymbol{k} = [100]^*$ propagation vector were observed below $T_N$ = 80 K and 22 K, respectively [Figures 8(a), 8(c), 8(e), S8, S9 and S10]; the temperature dependence of these peaks' intensities are shown in Figures 8(b), (d) and (f). This propagation vector corresponds to breaking of the lattice *C*-centring, leading to a primitive magnetic cell which is twice the size of the primitive nuclear cell.

Symmetry-mode analysis was used to determine the symmetry-allowed magnetic irreducible representations (irreps), yielding $mY_1^+$ and $mY_2^+$ irreps, in Miller and Love's notation,[70] for $Mn(NCS)_2$, $Fe(NCS)_2$ and $Co(NCS)_2$. Rietveld refinement of the data using these irreps revealed that only the $mY_2^+$ irrep was consistent with the data for $Mn(NCS)_2$, $Fe(NCS)_2$ and $Co(NCS)_2$. The refinements against the $mY_2^+$ symmetry-adapted mode yielded staggered magnetic moments of 4.02(4) $\mu_B$ ($Mn(NCS)_2$), 4.95(6) $\mu_B$ ($Fe(NCS)_2$) and 3.018(16) $\mu_B$ ($Co(NCS)_2$), with magnetic space group $P_A2_1/c$ in the BNS notation.[71]

In contrast to the other members of the $M(NCS)_2$ family, the magnetic Bragg peaks observed below $T_N$ = 56 K for $Ni(NCS)_2$ were indexed to a propagation vector of $\boldsymbol{k} = \left[ 00\tfrac{1}{2} \right]^*$, corresponding to $mA_1^+$ and $mA_2^+$ irreps. The neutron diffraction data were consistent with the $mA_2^+$-distorted structure and Rietveld refinement against this structure [Figures 8(g) and S11] yielded a staggered moment of 1.73(5) $\mu_B$ and a magnetic space group $C_c2/c$ in the BNS notation, with the magnetic unit cell doubled along the *c*-direction, relative to the nuclear cell [Figure 9].

Since $Ni(NCS)_2$ is ferromagnetic within the layer, we anticipate $J_1/J_2 > -0.5$ and $J_2 < 0$ (i.e. ferromagnetic), so that $Ni(NCS)_2$ lies in the FM region of the classical phase diagram [Figure 2(a)]. Likewise, we anticipate $J_1/J_2 < 0.5$ and $J_2 > 0$ for $Mn(NCS)_2$, $Fe(NCS)_2$ and $Co(NCS)_2$, so that these materials occupy the AF region of the classical phase diagram [Figure 2(a)]. Additional constraints on the relative sizes and signs of $J_1$ and $J_2$ are established in the Discussion.

The staggered moments of $Fe(NCS)_2$, $Co(NCS)_2$ and $Ni(NCS)_2$ are broadly consistent with the effective magnetic moments measured *via* bulk magnetic susceptibility measurements [Table 2], with deviations from the spin-only values likely arising from delocalisation (covalency) effects (see Discussion).

We also established the temperature dependence of the lattice parameters [Figure S12]. For all members of the $M(NCS)_2$ family, the lattice parameters were observed to vary broadly linearly with temperature, with no discontinuities observed. For $Mn(NCS)_2$, $Co(NCS)_2$ and $Ni(NCS)_2$, the *c*-axis expands most as temperature rises [Figure S12], since the interactions along the *c* axis are primarily weaker van der Waals interactions between the layers. The temperature dependence of lattice parameters for $Fe(NCS)_2$ could not be accurately determined, due to the breadth of the Bragg peaks (a consequence of the low crystallinity of the sample).

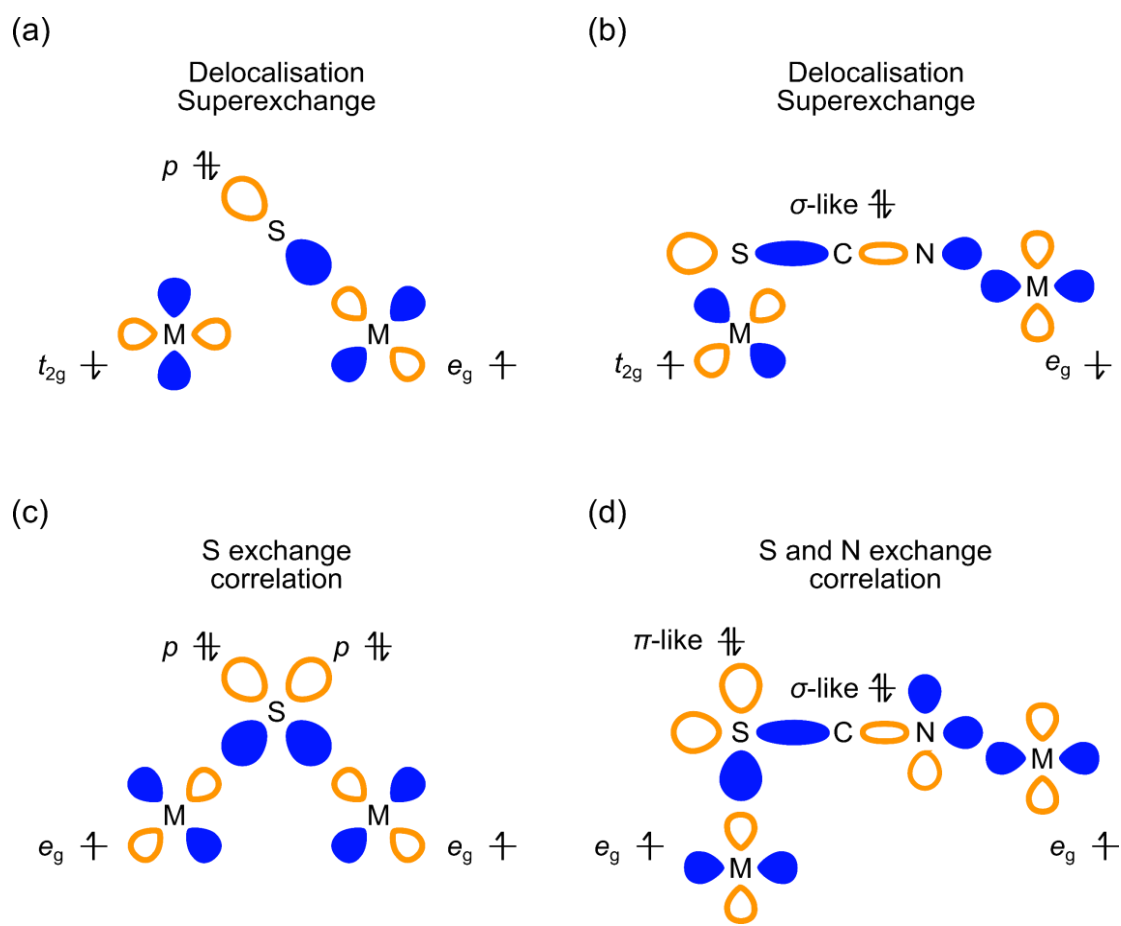

**Figure 10.** Schematic figures illustrating simplified superexchange mechanisms for the $J_1$ (**(a)** and **(c)**) and $J_2$ (**(b)** and **(d)**) interactions in terms of the magnetic $M^{2+}$ 3$d$ orbitals and frontier $\sigma$-like and $\pi$-like orbitals on (NCS)$^-$. Orange and blue lobes indicate opposite phases of orbital wavefunctions. The antiferromagnetic components of the $J_1$ (**(a)**) and $J_2$ (**(b)**) interactions proceed *via* delocalisation superexchange, whilst the ferromagnetic components proceed *via* exchange correlation—either *via* S only for $J_1$ (**(c)**) or *via* S and N for $J_2$ (**(d)**).

## 4. DISCUSSION

The divalent first-row TM thiocyanates $M$(NCS)$_2$ ($M$ = Mn, Fe, Co and Ni) adopt the same layered crystal structure, with space group $C2/m$. As $M$ changes, we observe broadly linear variations in the lattice parameters with the ionic radius of $M^{2+}$.

As earlier first-row TM ions occupy the $M^{2+}$ sites in the $M$(NCS)$_2$ structure, $\theta$ becomes increasingly antiferromagnetic [Figure 7], due to (individually) stronger antiferromagnetic interactions and/or weaker ferromagnetic interactions. We expect that the magnetic properties of the $M$(NCS)$_2$ lattices are dominated by the two nearest neighbour superexchange interactions, $J_1$ and $J_2$. The dipolar interactions between $M^{2+}$ are expected to be small, due to the large distances between $M^{2+}$ centres (at least 3.6 Å, corresponding to a dipolar interaction strength on the order of 0.01 K); likewise, we expect direct exchange to be weak, due to the large $M^{2+}$–$M^{2+}$ distances. The interlayer interaction, $J_3$, will also likely be significantly smaller than the intralayer interactions, due to the large separation and lack of chemical bonds between $M^{2+}$ ions. The observed interlayer S⋯S contacts are comparable to those of other van der Waals magnetic materials which are well-described by low-dimensional Hamiltonians.[32–35,43–45]

Therefore, we expect the strengths of the dominant exchange interactions in $M$(NCS)$_2$ ($J_1$ and $J_2$) to depend primarily on the occupation of the magnetic 3$d$ orbitals and their overlap with the frontier orbitals on (NCS)$^-$. In all cases, we expect the anti-bonding orbitals generated from this overlap to be the highest occupied orbitals and therefore these will be the orbitals which dominate the magnetic exchange interactions in $M$(NCS)$_2$.[72]

The $J_1$ interaction will likely contain two contributions: firstly, antiferromagnetic superexchange involving overlap between $t_{2g}$ and $e_g$ orbitals on adjacent $M^{2+}$ ions (mediated by a S $p$-like orbital, Figure 10(a)) and secondly, ferromagnetic exchange-correlation-driven superexchange mediated by the two orthogonal sulphur $p$-like orbitals [Figure 10(c)].[73–75] The antiferromagnetic component will weaken from left to right across the TM series, as the number of unpaired electrons in $t_{2g}$ orbitals decreases, until for $M$ = Ni, the $t_{2g}$ orbitals are fully occupied and we anticipate a negligible antiferromagnetic component. Each metal has the same $e_g$ orbital occupation ($e_g^2$) and so we expect the ferromagnetic contribution [Figure 10(c)] to be broadly similar in magnitude across the series. Consequently, we expect the $J_1$ interaction to become increasingly ferromagnetic (more negative $J_1$) for later TM.

The $J_2$ superexchange interaction is principally mediated via the $\sigma$- and $\pi$-like frontier orbitals on (NCS)$^-$, dominated by lobes on S and N atoms [Figure 10 (b) and (d)].[9] An antiferromagnetic contribution will arise from delocalisation superexchange between metal $t_{2g}$ and $e_g$ orbitals via the $\sigma$-like frontier orbital on (NCS)$^-$ [Figure 10(b)]. A countervailing ferromagnetic interaction will be produced by exchange correlation between the metal $e_g$ orbitals on $M^{2+}$, mediated by the orthogonal $\sigma$- and $\pi$-like frontier orbitals on (NCS)$^-$ [Figure 10(d)]. Again, as electrons are added to the $t_{2g}$ orbitals (from Mn$^{2+}$ to Ni$^{2+}$), the antiferromagnetic component of the $J_2$ interaction will weaken and become negligible for $M$ = Ni, resulting in a more ferromagnetic $J_2$. Indeed, for many $M$(NCS)$_2$ solvate frameworks (those of the form $M$(NCS)$_2L_x$, where $L$ = ligand), the Ni member of the family orders ferromagnetically along the Ni–NCS–Ni chains, whilst the Mn and Co members order antiferromagnetically,[23,25,60,76,77] consistent with our experimental results and proposed magnetochemical mechanisms.

This rationalisation can only provide a general understanding of the magnetic behaviour of these materials. Deviations from the 'ideal' 90° $M$–S–$M$ and 180° $M$–(NCS)–$M$ bond angles, will likely mix the ferromagnetic and antiferromagnetic components of each interaction, giving deviations from the trends we explain above. In addition, changes in the size and energy match of the metal 3$d$ and (NCS)$^-$ orbitals, along with spin-orbit effects, will play a key role in determining the strength of the interactions in these materials.

Furthermore, we expect that single-ion properties will play a significant role in the behaviour of Co(NCS)$_2$, Ni(NCS)$_2$ and Fe(NCS)$_2$, producing deviations both from classical two-dimensional Heisenberg behaviour and in the measured value of $\theta$. Nevertheless, our qualitative predictions of the signs and strengths of the $J_1$ and $J_2$ interactions are borne out by the observed magnetic ground states and trends in the bulk magnetic properties.

We anticipate that these materials will show similar magnetic behaviour to two-dimensional systems, despite ordering as bulk antiferromagnets, as we expect the interlayer interactions ($J_3$) to be weak. Therefore, our magnetochemical model allows us to rationalise the location of each compound in the classical two-dimensional magnetic phase diagram [Figure 2(a)]. For $M$ = Mn, Fe and Co, we expect $J_2 > 0$ (i.e. antiferromagnetic), whilst we expect $J_1$ to be small and either ferromagnetic or antiferromagnetic, due to cancellation of the superexchange contributions. This predicts that these materials all lie in the AF region of the phase diagram, as observed from their ground state magnetic structures. As $J_1$ becomes more antiferromagnetic, we anticipate moderate geometric frustration with the antiferromagnetic $J_2$, and indeed $f$ is largest for Mn(NCS)$_2$. For $M$ = Ni, we anticipate $J_1$ and $J_2$ to be ferromagnetic (i.e. $J_1, J_2 < 0$), placing it in the FM region of the phase diagram, as observed.

For all members of the $M$(NCS)$_2$ family, the size of the staggered moments obtained from PND are all lower than the size of the expected maximum ordered moment (i.e. $gS$). This likely arises from the effects of covalency, where spin density is transferred from $M^{2+}$ to the (NCS)$^-$ ligand,[78] and from contributions from the orbital moment.

Finally, we find that the ordered moment lies broadly along the N–$M$–N axis for Mn(NCS)$_2$ and Ni(NCS)$_2$, suggesting significant spin density in the $d$ orbital pointing along that axis. For Fe(NCS)$_2$ and Co(NCS)$_2$, the moment has a significant tilt away from the N–$M$–N axis, towards the meridional plane containing S-bound (NCS)$^-$ ligands, suggesting the important role of spin-orbit effects. Future electron spin resonance spectroscopy will shed further light on the origin of these differences by providing more precise estimates of $g$ and the superexchange interactions, respectively.

Recent work on $CrI_3$ and $Cr_2Ge_2Te_6$ has shown that mechanical exfoliation of layered ferromagnets can generate single-layer magnets.[79–81] While bulk $Ni(NCS)_2$ orders as a three-dimensional antiferromagnet, the combination of ferromagnetic intralayer order with weak interlayer interactions suggests that monolayers of this may host single-layer ferromagnetism, provided there is sufficient anisotropy. These single-layer ferromagnets may have applications as magnetoelectronic devices, ferromagnetic light emitters[79,80] and hybrid multilayer materials,[82,83] motivating future synthetic studies.

## 5. CONCLUSION

In this work, we have determined how the strengths of magnetic interactions vary across the first-row TM pseudobinary thiocyanates, $M(NCS)_2$. Two new materials—$Mn(NCS)_2$ and $Fe(NCS)_2$—are reported, alongside the magnetic structures of these materials and of $Co(NCS)_2$ and $Ni(NCS)_2$. Based on the observed magnetic structures, we have qualitatively rationalised the relative strengths and signs of the nearest-neighbour in-plane exchange interactions, $J_1$ and $J_2$, and have located each material on the magnetic phase diagram of the spatially anisotropic triangular lattice.

On moving to TM ions earlier in the row, the net magnetic interactions become stronger and increasingly antiferromagnetic, with the Weiss constant reaching −115(3) K for $Mn(NCS)_2$ and the Néel temperature of $Fe(NCS)_2$ reaching above 78 K.

PND revealed that $Mn(NCS)_2$, $Fe(NCS)_2$ and $Co(NCS)_2$ adopt the same commensurate antiferromagnetic structure, in which parallel spins along the crystallographic $b$ axis order antiferromagnetically along the $a$ axis; these layers of spins are then stacked antiferromagnetically along the $c$-direction. In contrast, the magnetic structure of $Ni(NCS)_2$ comprises ferromagnetically ordered $ab$ layers ordered antiferromagnetically along the $c$-axis. This suggests that single-layer $Ni(NCS)_2$ may be a candidate monolayer ferromagnet belonging to a new family of magnetic frameworks. The results collected from this study open up new avenues for the rational design of magnetic molecular framework materials.

## AUTHOR INFORMATION

**Corresponding Authors**

*E-mail: cpg27@cam.ac.uk; matthew.cliffe@nottingham.ac.uk

**ORCID**

Euan N. Bassey: 0000-0001-8827-7175
Joseph A.M. Paddison: 0000-0002-2274-0988
Evan N. Keyzer: 0000-0002-0655-2813
Jeongjae Lee: 0000-0003-4294-4993
Pascal Manuel: 0000-0002-8845-6576
Ivan da Silva: 0000-0002-4472-9675
Siân E. Dutton: 0000-0003-0984-5504
Clare P. Grey: 0000-0001-5572-192X
Matthew J. Cliffe: 0000-0002-0408-7647

**Notes**

There are no conflicts of interest to declare.

## ACKNOWLEDGEMENTS

E.N.B. thanks the EPSRC for financial support. J.A.M.P.'s work at Cambridge was supported by Churchill College, University of Cambridge. J.A.M.P.'s work at Oak Ridge National Laboratory (ORNL) was supported by the Laboratory Directed Research and Development Program of ORNL, managed by UT-Battelle, LLC for the US Department of Energy (discussion of magnetic modelling) and the U.S. Department of Energy, Office of Science, Basic Energy Sciences, Materials Sciences and Engineering Division (computational resources). E.N.K. thanks NSERC of Canada for a PGSD. J.L. thanks Trinity College, University of Cambridge for financial support. M.J.C. acknowledges the School of Chemistry, University of Nottingham for a Hobday Fellowship. Magnetic measurements were carried out using the Advanced Materials Characterisation Suite, funded by EPSRC Strategic Equipment Grant EP/M000524/1. We also acknowledge the Rutherford Appleton Laboratory for access to the ISIS Neutron Source.

## Supporting Information

The Supporting Information is available free of charge on the [ACS Publications website](ACS Publications website) at DOI:

Calculated structural coordinates in CIF format

Additional experimental details, including powder X-ray diffraction, TGA, isothermal magnetisation, powder neutron diffraction and variable temperature powder neutron diffraction data.

For Table of Contents Only

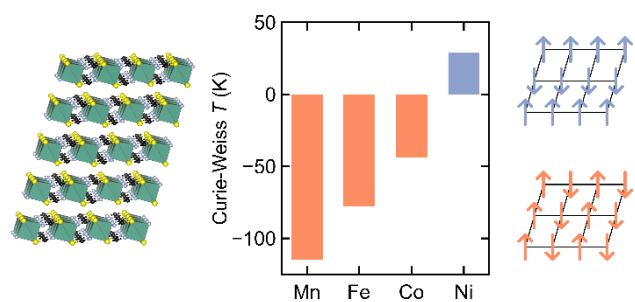

Synopsis:

We report the synthesis, properties and magnetic structures of $M$(NCS)$_2$ ($M$ = Mn, Fe, Co, Ni) and find that as earlier transition metals are used, the net magnetic interaction strength increases. Mn(NCS)$_2$, Fe(NCS)$_2$ and Co(NCS)$_2$ show stripe-antiferromagnetic order, but Ni(NCS)$_2$ contains ferromagnetic layers, coupled antiferromagnetically.